\documentclass[aps,prc,superscriptaddress,twocolumn,amsfonts,amsmath,amssymb,showpacs,floatfix]{revtex4-1}

\usepackage{graphicx}
\usepackage{longtable}
\usepackage{bm}
\usepackage{multirow}
\usepackage{hyperref}
\usepackage{ulem}
\hypersetup{colorlinks,citecolor=blue,filecolor=blue,linkcolor=blue,urlcolor=blue}
\usepackage{color}
\usepackage{dcolumn}
\usepackage{lipsum}

\begin{document}

\title{Consistency of nucleon-transfer sum rules in well-deformed nuclei}

\author{B.~P.~Kay}
\email[E-mail: ]{kay@anl.gov}
\affiliation{Physics Division, Argonne National Laboratory, Argonne, Illinois 60439, USA}
\author{J.~P.~Schiffer}
\affiliation{Physics Division, Argonne National Laboratory, Argonne, Illinois 60439, USA}
\author{S.~J.~Freeman}
\affiliation{Department of Physics and Astronomy, The University of Manchester, Oxford Road, Manchester, M13 9PL, UK}
\author{T.~L.~Tang}
\affiliation{Physics Division, Argonne National Laboratory, Argonne, Illinois 60439, USA}
\author{B.~D.~Cropper}
\affiliation{Department of Physics and Astronomy, The University of Manchester, Oxford Road, Manchester, M13 9PL, UK}
\author{T.~Faestermann}
\affiliation{Physik Department E12, Technische Universit\"at M\"unchen, D-85748 Garching, Germany}
\affiliation{Maier-Leibnitz Laboratorium der M\"{u}nchner Universit\"{a}ten, D-85748 Garching, Germany}
\author{R.~Hertenberger}
\affiliation{Fakult\"at f\"ur Physik, Ludwig-Maximilians Universit\"at M\"unchen, D-85748 Garching, Germany}
\author{J.~M.~Keatings}
\affiliation{SUPA, School of Computing, Engineering, and Physical Sciences, University of the West of Scotland,
Paisley PA1 2BE, United Kingdom}
\author{P.~T.~MacGregor}
\affiliation{Department of Physics and Astronomy, The University of Manchester, Oxford Road, Manchester, M13 9PL, UK}
\author{J.~F.~Smith}
\affiliation{SUPA, School of Computing, Engineering, and Physical Sciences, University of the West of Scotland,
Paisley PA1 2BE, United Kingdom}
\author{H.-F.~Wirth}
\affiliation{Fakult\"at f\"ur Physik, Ludwig-Maximilians Universit\"at M\"unchen, D-85748 Garching, Germany}


\begin{abstract}

Nucleon-transfer sum rules have been assessed via a consistent reanalysis of cross-section data from neutron-adding ($d$,$p$) and -removing ($d$,$t$) reactions on well-deformed isotopes of Gd, Dy, Er, Yb, and W, with $92\leq N\leq108$, studied at the Niels Bohr Institute in the 1960s and 1970s. These are complemented by new measurements of cross sections using the ($d$,$p$), ($d$,$t$), and ($p$,$d$) reactions on a subset of these nuclei. The sum rules, defined in a Nilsson-model framework, are remarkably consistent. A single overall normalization is used in the analysis, which appears to be sensitive to assumptions about the reaction mechanism, and in the case of sums using the ($d$,$t$) reaction, differs from values determined from reactions on spherical systems. 

\end{abstract}

\pacs{}

\maketitle

\section{Introduction}

Nucleon-transfer sum rules~\cite{Macfarlane60} have been explored at length over the last 60 years since they were proposed, perhaps most extensively in recent surveys of closed-shell and other near-spherical \mbox{systems~\cite{Schiffer12,Schiffer13,Kay13}}. The sum rules state that the summed spectroscopic strength from nucleon-adding and -removing reactions to final states based on an orbital $j$ equals the total $(2j+1)N_j$ degeneracy of the orbital. $N_j$ is a normalization factor and should be equal to 1.0 if the independent single-particle model of nuclear structure and the description of the reaction mechanism are correct. A consequence of short-range correlations between nucleons means that the model description accounts for only about 60\% of the total single-particle strength. This is supported by a wealth of reaction data, largely on stable nuclei, using probes such as ($e$,$e'p$), ($p$,$2p$), and nucleon-transfer reactions, where $N_j$ is consistently found to be about 0.6~\cite{Kramer01,Kay13}. This reduction from unity is often called {\it quenching} and is seemingly independent of mass, reaction type, and angular momentum.

In regions of the chart of nuclides where nuclei have permanent quadrupole deformations, spherical single-particle symmetry is broken and a given (2$j+1$)-fold degenerate shell-model orbital is fragmented into $j+1/2$ ($\Omega=1/2$, 3/2, ..., $j$) Nilsson states~\cite{Nilsson55,Nilsson95}. These Nilsson states correspond to different projections of the motion on the deformation axis, each having a degeneracy of two nucleons. 

For a given shell-model orbital, the strength is distributed among the deformed states according to the Nilsson coefficients $C^2_{jl}$. Each Nilsson state in an odd-A nucleus has a rotational band built on it, and the appropriate pattern of spectroscopic factors yields a distinctive fingerprint of transfer strength over the rotational band members. Expressions for reaction probabilities to states in deformed nuclei were first deduced by Satchler in the late 1950s~\cite{Satchler55,Satchler58}---and are summarized and discussed with reference to experimental results by Elbek and Tj\o m~\cite{Elbek69}. The relevant details are summarized here. The nucleon-adding and -removing cross sections on an even-even target to a state $j$ can be expressed as
\begin{equation}\label{eqn1}
\sigma^{\pm}_{\rm exp}=2N_jg^{\pm} \left[C_{j\ell}P_j\right]^2 \sigma^{\pm}_{\rm DWBA},
\end{equation}
 where $\sigma_{\rm exp}$ and $\sigma_{\rm DWBA}$ are the experimental and calculated cross sections; the latter typically determined using the distorted wave Born approximation (DWBA). $g^{\pm}$ are the reciprocal of statistical factors. The $P^2$ term is $V^2$ for ($-$) removing reactions, the degree to which a pair of nucleons occupy the orbital, and $U^2$ for ($+$) adding reactions, the degree to which they do not, and $U^2+V^2=1$. Here, $C_{j\ell}$ is the Nilsson coefficient. Coriolis effects, among others, can cause mixing. The amplitude of this mixing, $\alpha_i$, can be calculated and Eq.~\ref{eqn1} can be modified accordingly by replacing the term in brackets by $\left[\sum_i\alpha_i C^i_{j\ell}P_j^i\right]^2$, where the sum is over the admixed bands (see, for example, Ref.~\cite{Casten00}). 

Several sum rules apply to transfer reactions in the Nilsson description. By considering a fixed $j$ and Nilsson quantum numbers $\Omega^{\pi}[Nn_z\Lambda]$~\cite{Elbek69} a sum rule can be stated making use of the fact that only one pair of nucleons can have those quantum numbers. For transfer reactions on an even-even nucleus, in the absence of mixing, this is the sum of the strength determined from a single state, $j^\pi$, in each of the adding and removal reactions, which must equal the degeneracy of the orbital, i.e., 2. From Eq.~\ref{eqn1}, it follows that
\clearpage
\begin{equation}\label{eqn2}
\frac{1}{C_{j\ell}^2}\bigg(\frac{(2j+1)\sigma^+_{\rm exp}}{\sigma^+_{\rm DWBA}}+\frac{\sigma^-_{\rm exp}}{\sigma^-_{\rm DWBA}}\bigg)=2N_j.
\end{equation}

For given values of $j^{\pi}$ and $\Omega^{\pi}[Nn_z\Lambda]$, it is possible to track a given excitation across several chains of isotopes and assess the degree to which the analysis yields consistent sums, and how much the value differs from the expected value of two, i.e., what is the value of $N_j$. To the best of our knowledge, such a study has not been carried out before and past analyses have not been carried out with modern finite-range DWBA codes and optical-model parameterizations. 

\section{The Niels Bohr Institute Data}

\subsection{The cross sections}

To explore this nucleon-transfer sum rule, data collected at the Niels Bohr Institute (NBI) in the late 1960s and early 1970s by Elbek and collaborators~\cite{Tjom67,Grotdal70,Tjom69,Burke66,Casten72} were initially used. The ($d$,$p$) and ($d$,$t$) reaction cross sections  were measured simultaneously using a broad-range (large-acceptance) magnetic spectrograph~\cite{Borggreen63} with photographic plates at the focal plane. The incident deuteron energies for all of the studies discussed here were close to 12~MeV, and the $Q$-value resolution was around 6-12~keV FWHM. The uncertainties on the reported cross sections were estimated to be around 10-15\% for relative cross sections, that is for peaks within each spectrum, and around 15-20\% in absolute terms for cross sections larger than 20~$\mu$b/sr.

The cross sections were recorded at angles of 60$^{\circ}$, 90$^{\circ}$, and 125$^{\circ}$, with some exceptions where contaminants obscured peaks. At these energies, the ($d$,$p$) and ($d$,$t$) angular distributions have broad maxima between $\theta_{\rm lab}=60^{\circ}$ to 90$^{\circ}$. In most cases $j^{\pi}$ assignments were already known at the time that these measurements were carried out, so detailed angular distributions were not required for spin assignments. The momentum matching in the ($d$,$p$) and ($d$,$t$) reactions is well suited for $\ell=1$ transfer and has proven to be reliable for $\ell=3$ transfer~\cite{Schiffer13}. We note that for all isotopes studied here, the outgoing triton in the ($d$,$t$) reaction was within an MeV or so of the Coulomb barrier. 

The data collected were on the ($d$,$p$) and ($d$,$t$) reactions on stable, even isotopes of Gd ($Z=64$)~\cite{Tjom67}, Dy ($Z=66$)~\cite{Grotdal70}, Er ($Z=68$)~\cite{Tjom69}, Yb ($Z=70$)~\cite{Burke66}, and W ($Z=74$)~\cite{Casten72}, spanning a range in neutron number from 92 to 108. These nuclei all have ground-state deformation parameters $\beta$ greater than 0.3 (with the exception of the $^{180,182}$W, which are around 0.25)~\cite{Raman01} and all have ratios of $E(4^+)$ to $E(2^+)$ of $\sim$3.3~\cite{ensdf}, i.e., they have well-defined rotational ground-state bands. A list of the isotopes and some of their relevant properties are given in Table~\ref{tab1}.

\subsection{Analysis of the Niels Bohr Institute data}

The sum of the strengths (Eq.~\ref{eqn2}) for the adding and removing reactions on a given target for the $j^{\pi}=1/2^-$ member of the $1/2^-[521]$ band can be determined from these data for fifteen cases. The summed strength of the $j^{\pi}=7/2^-$ member of the $5/2^-[512]$ band was similarly extracted in three isotopes of Er and and three of Yb. The $C^2_{j\ell}$ coefficients were calculated using the method described in Chi~\cite{Chi66} and are listed in Table~\ref{tab1}. Note that the $C^2_{j\ell}$ coefficients for the two different bands are markedly different.

To calculate the DWBA cross sections, a standard approach was taken, as described, for instance in  \mbox{Refs.~\cite{Schiffer12,Schiffer13,Kay13}}. The finite-range distorted-wave-Born-approximation code {\sc ptolemy}~\cite{ptolemy} was used throughout. For the ($d$,$p$) reaction, the deuteron wave function was described by the Argonne $\nu_{18}$ potential~\cite{Wiringa95}.  The target bound-state wave functions were defined by a Woods-Saxon potential with a spin-orbit derivative term, having parameters $r_0=1.28$~fm, $a=0.65$~fm, $V_{\rm so}=6$~MeV, $r_{\rm so0}=1.1$~fm, and $a_{so}=0.65$~fm. In the case of the ($d$,$t$) reactions, the parameterized potential of Brida {\it et al.}~\cite{Brida11} defined the wave functions and the also the $\langle d|t\rangle$ overlap. No allowance is  made for deformation in the DWBA calculations.

The global optical-model potentials of An and Cai ~\cite{An06} were used for the deuterons, those of Koning and Delaroche~\cite{Koning03} for the protons, and those of Pang {\it et al.}~\cite{Pang09} for the tritons. All of these have been assessed on a broad range of nuclei and have led to largely consistent results for spherical nuclei~\cite{Kay13}, and include in the data from which they are derived, nuclei close to, but not in, the rare-earth region. It is not clear how well these global parameterizations can approximate the distorted wave functions in the incident and outgoing channels, those of the bound-states of the transferred neutrons and the overlaps between them in deformed nuclei. Coupled-channel effects, that account for some of these, are expected to be important in deformed nuclei, with strongly enhanced $B(E2)$ values from 0$^+$ ground states to at least the first (2$^+$) state in the rotational band, and similar couplings in odd-A nuclei. There have been several previous studies of these effects. Both modifications to the shapes of the angular distributions can be seen, as well as the magnitude of the cross section~\cite{GlendenningBook} at the 10-20\% level depending on the case~\cite{Siemssen67,Kunz69}. The work of Kunz, Rost, and Johnson~\cite{Kunz69} finds that in a DWBA analysis, coupled-channel effects may be approximated by a larger bound state radius which leads to larger cross sections. 

\begin{table*}
\caption{\label{tab1} The normalized strength for neutron adding and removing and the summed strength from the present analysis, derived from the NBI ($d$,$p$) and ($d$,$t$) reaction data. The average normalization factor is $\overline{N}=1.18(15)$. The uncertainties on the sums are discussed in the text. The $E_{4^+_1}/E_{2^+_1}$ ratio is given, along with the ground-state deformation parameter $\beta$, the orbital label ($j^{\pi}$, $\Omega^{\pi}[Nn_z\Lambda]$), and the Nilsson coefficients ($C_{j\ell}$).}
\newcommand\T{\rule{0pt}{3ex}}
\newcommand \B{\rule[-1.8ex]{0pt}{0pt}}
\begin{ruledtabular}
\begin{tabular}{cccccccccc}
$j^{\pi}$, $\Omega^{\pi}[Nn_z\Lambda]$ & Isotope \B\B & $Z$ & $N$ & $E_{4^+_1}/E_{2^+_1}$ &  $\beta$\footnote{The quoted ground-state deformation parameters, $\beta$, are taken from Ref.~\cite{Raman01} and defined as $\beta=(4\pi/3ZR_0^2)[B(E2)\uparrow/e^2]^{1/2}$.} &   $C^2_{j\ell}$\footnote{The $C_{j\ell}$ coefficients calculated from the formalism in Ref.~\cite{Chi66} using the code of Ref.~\cite{Tang19}.} & Adding & Removing & Sum  \\
\hline
$\frac{1}{2}^-$, $\frac{1}{2}^-[521]$ & $^{160}$Gd\T	&	64	&	96	&	3.302	&	0.353	&		0.245	& 1.21 & 0.46 &		1.67	\\
& $^{158}$Dy\T	&	66	&	92	&	3.206	&	0.326	&			0.248	&	 1.46 &	0.52 &	1.98	\\
& $^{160}$Dy\T	&	66	&	94	&	3.270	&	0.339	&			0.247	&	 1.20 &	0.39 &	1.59	\\
& $^{162}$Dy\T	&	66	&	96	&	3.294	&	0.343	&			0.246	&	 1.04 &	0.76 &	1.81	\\
& $^{164}$Dy\T	&	66	&	98	&	3.301	&	0.348	&			0.246	&	 1.31 &	0.43 &	1.73	\\
& $^{164}$Er\T	&	68	&	96	&	3.277	&	0.333	&			0.247	&	 1.27 &	0.58 &	1.86	\\
& $^{166}$Er\T	&	68	&	98	&	3.289	&	0.342	&			0.246	&	 1.50 &	0.81 &	2.31	\\
& $^{168}$Er\T	&	68	&	100	&	3.309	&	0.338	&			0.247	&	 1.45 &	1.12 &	2.58	\\
& $^{170}$Er\T	&	68	&	102	&	3.310	&	0.336	&			0.247	&	 0.49 &	1.73 &	2.22	\\
& $^{168}$Yb\T	&	70	&	98	&	3.266	&	0.322	&			0.248	&	 1.33 &	0.85 &	2.18	\\
& $^{170}$Yb\T	&	70	&	100	&	3.293	&	0.326	&			0.248	&	 0.78 &	1.18 &	1.96	\\
& $^{172}$Yb\T	&	70	&	102	&	3.305	&	0.330	&			0.247	&	 0.47 &	1.93 &	2.41	\\
& $^{174}$Yb\T	&	70	&	104	&	3.310	&	0.325	&			0.248	&	 0.12 &	1.94 &	2.06	\\
& $^{180}$W\T	&	74	&	106	&	3.260	&	0.254	&			0.255	&	 0.14 &	1.70 &	1.83	\\
& $^{182}$W\T\B	&	74	&	108	&	3.291	&	0.251	&			0.256	 & 0.09 &	  1.91 &	2.00	\\
\hline
$\frac{7}{2}^-$, $\frac{5}{2}^-[512]$	& $^{166}$Er\T	&	68	&	98	&	3.289	&	0.342	& 	0.783	&	1.67  &	0.15 & 	1.82	\\
& $^{168}$Er\T	&	68	&	100	&	3.309	&	0.338	&			0.784	&	 1.78  &	0.40 &	2.18	\\. 
& $^{170}$Er\T	&	68	&	102	&	3.310	&	0.336	&			0.784	&	  1.24 &	0.84 &	2.08	\\
& $^{170}$Yb\T	&	70	&	100	&	3.293	&	0.326	&			0.785	&	 1.19  &	0.46 &	1.65	\\
& $^{172}$Yb\T	&	70	&	102	&	3.305	&	0.330	&			0.785	&	  1.23 &	0.66 &	1.89	\\
& $^{174}$Yb\T	&	70	&	104	&	3.310	&	0.325	&			0.785	&	  0.46 &	1.72 &	2.18	\\
\end{tabular}
\end{ruledtabular}
\end{table*}

The $1/2^-[521]$ band can be considered a so-called decoupled band, being at the rotationally-aligned coupling limit as evidenced by the energy spacing of its \mbox{$j$, $j+2$, ...,} members matching the energy spacing of the even-even core \mbox{0$^+$, 2$^+$, ...,} excitations~\cite{Stephens75}. This is the case of all nuclei studied here. For this $\Omega=1/2$ case, a decoupling parameter $a_d$ modifies the $C_{j\ell}$ coefficient, such that $a_d=-\sum_{j=1/2}^{N+1/2}(-)^{j+1/2}(j+1/2)C^2_{j\ell}$. For the $1/2^-[521]$ bands studied here, $a_d$ ranges from 0.86 to 0.93. These corrections are assumed negligible. For the $5/2^-[512]$ band, it lies neither in the rotationally-aligned limit nor the strongly-coupled limit with an odd-$A$-to-average-even-even transition energy ratio of around 2. In this analysis, Coriolis mixing matrix elements have not been included. Coriolis mixing will modify the cross sections. Previous analyses of data for several isotopes of Gd, Dy, Er, and Yb, suggest these modifications are small, around 20\%~\cite{Kanestrom69}. A detailed exploration of Coriolis couplings can be found in Ref.~\cite{Casten72}, which is part of the original analysis of the tungsten data used here. Note that Coriolis couplings are negligible for the $1/2^-[521]$ band, which is used in the bulk of our reanalysis, and only non-negligible (10-20\%) for the $5/2^-[512]$ band. It is also assumed that other forms of coupling/mixing are small.

Table~\ref{tab1} shows the normalized summed spectroscopic factors derived from this analysis. An average normalization, $\overline{N}$, derived from 21 different determinations of $N_j$ has been deduced (Eqn.~\ref{eqn2}), yielding $\overline{N}=1.18(15)$ across all isotopes. The rms spread in the sums is 15\%, which is similar to the estimated relative uncertainty in the cross sections. Similar values are found for subsets of the data, with the $\ell=1$ transfer giving $\overline{N}_{1/2}=1.19(17)$ and the $\ell=3$ transfer, $\overline{N}_{7/2}=1.16(13)$. 

\begin{figure*}
\includegraphics[scale=0.8]{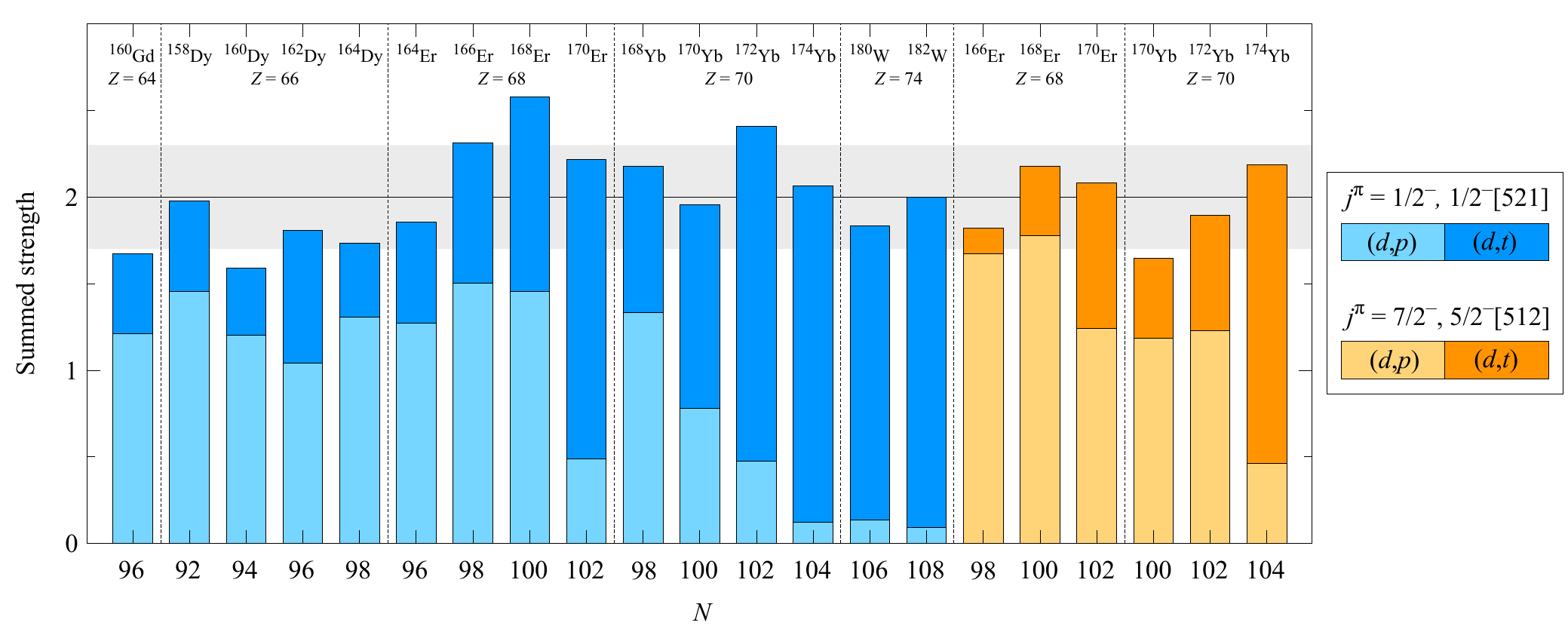}
\caption{\label{fig1} The normalized summed strength, defined in the text, for Nilsson orbitals as determined from ($d$,$p$) [lighter shade] and ($d$,$t$) [darker shade] reaction data from NBI on isotopes of Gd~\cite{Tjom67}, Dy~\cite{Grotdal70}, Er~\cite{Tjom69}, Yb~\cite{Burke66}, and W~\cite{Casten72}.  The blue bars (left) are for the $j^{\pi}=1/2^-$ member of the $1/2^-[521]$ band and the orange bars (right) are for the $j^{\pi}=7/2^-$ member of $5/2^-[512]$ band. The grey band is the rms spread.}
\end{figure*}

\begin{figure}
\includegraphics[scale=0.84]{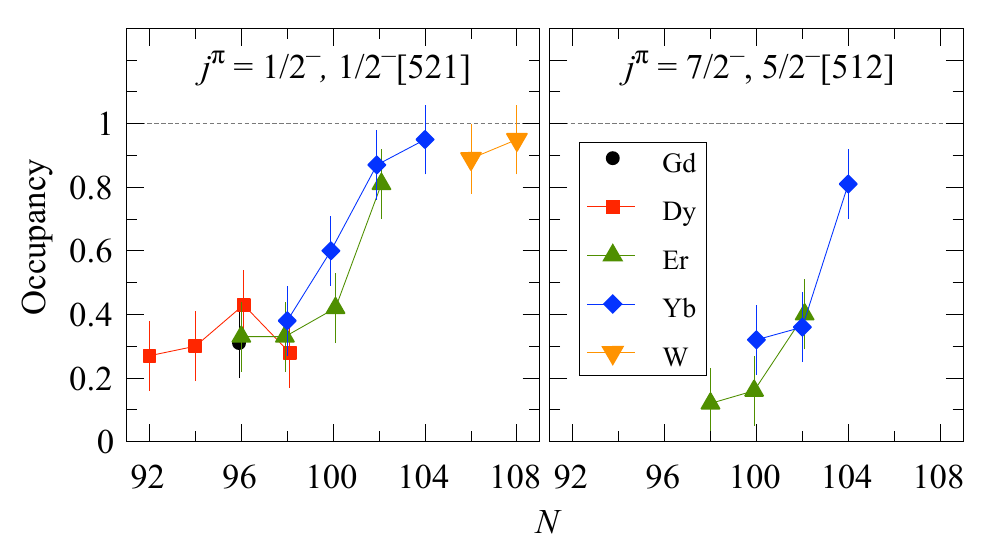}
\caption{\label{fig2} The occupancy, derived from the average of the adding and removing strength as determined from the NBI data~\cite{Tjom67,Grotdal70,Tjom69,Burke66,Casten72}, as a function of neutron number, $N$.}
\end{figure}

This consistency is quite remarkable across $\ell=1$ and 3 transfer, for final states with significantly different $C^2_{j\ell}$ coefficients and occupancies, from $64\leq Z\leq74$ and $96\leq N\leq108$. It shows that the sum rules, expressed in this manner (Eqn.~\ref{eqn2}) in the framework of the Nilsson model, are robust for these well-deformed nuclei. Figure~\ref{fig1}, shows the normalized summed strength and the fraction of which is the ($d$,$p$) and ($d$,$t$) strength. The emptying ($U^2$) and filling ($V^2$) of pairs of nucleons in these orbitals can be seen, most dramatically in the Yb isotopes between $98\leq N\leq104$. The occupancy, $V^2$, is shown in Fig.~\ref{fig2} to emphasize the filling pattern, where $V^2$ is the derived from $U^2+V^2=1$.

The value of the normalization is quite a bit larger than the 0.6 found in spherical nuclei. It was already noted by Erskine and Siemssen~\cite{Siemssen67} in their study of tungsten that the spectroscopic factors were larger than in spherical systems, and the reason for this was somewhat explained by Kunz, Rost, and Johnson~\cite{Kunz69} who pointed out the need for a larger radius for the bound state, to account for the effect of deformation on the potentials. In exploring the sensitivity of DWBA calculated cross sections, albeit it in a spherical basis, we note that the ($d$,$t$) cross sections are much more sensitive, by as much as a factor of five, to the radius of the bound state wave functions than for the ($d$,$p$) and ($p$,$d$) reactions. This is perhaps because of the much stronger surface absorption in the distorting parameters for tritons than those for protons and deuterons. 

\section{The Munich Data and analysis }

To further explore these systems, and as an independent test of the NBI absolute cross sections, we collected new data using the tandem facility at Munich. The ($d$,$p$), ($d$,$t$), and ($p$,$d$) reactions were carried out on targets of $^{170}$Er and $^{174}$Yb. The ($d$,$p$) and ($d$,$t$) reactions were carried out at the same energy of 12~MeV as the NBI measurements, and the ($p$,$d$) reaction at 18~MeV. As noted above, DWBA cross sections for the ($d$,$t$) reaction are more sensitive to choices of bound-state and optical-model potentials than the ($p$,$d$) reaction, so measurements with the latter reaction might reveal some further insights into the reaction mechanism and analyses. The ($d$,$p$) and ($p$,$d$) reactions were carried out at energies ideal for $\ell=1$ and 3 transfer, a few MeV/u above the Coulomb barrier in the entrance and exit channels~\cite{Schiffer12,Schiffer13}, in contrast to the ($d$,$t$) reaction where the tritons are closer to the barrier in the exit channel.

The data were taken and cross sections deduced in the same manner as those described in detail in Ref.~\cite{Freeman17}. Beams of 12-MeV deuterons and 18-MeV protons were delivered from the MP tandem accelerator at the Maier-Leibnitz Laboratorium (MLL) to the Q3D spectrometer~\cite{Scheerer76}, which was used to momentum analyze the reaction products. 

\begin{figure}
\includegraphics[scale=0.84]{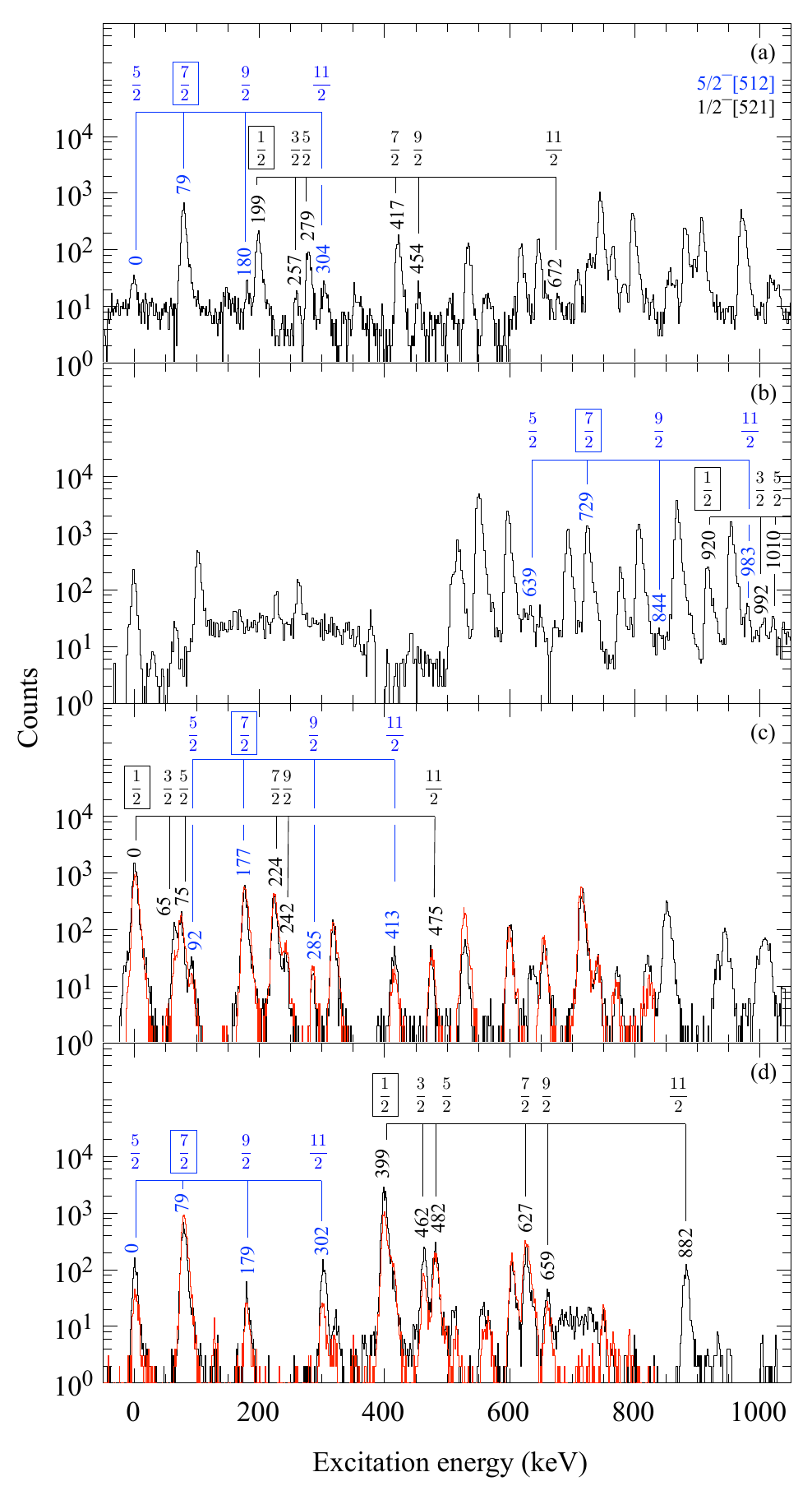}
\caption{\label{fig3} (a) Excitation-energy spectra for the $^{170}$Er($d$,$p$)$^{171}$Er reaction at 12 MeV for $\theta_{\rm lab}=90^{\circ}$. States belonging to the 5/2$^-$[512] (blue lines, text) and 1/2$^-$[521] (black lines, text) Nilsson configurations are labeled by their energy and spin. The states with their spin value given in a rectangle are those of interest in this work. The lines connect the states in a given band, highlighting their similarities in each isotope. (b) The same as (a) for the $^{174}$Yb($d$,$p$)$^{175}$Yb reaction at 12 MeV for $\theta_{\rm lab}=90^{\circ}$. (c) Shows yields for both the $^{170}$Er($p$,$d$)$^{169}$Er reaction at 18 MeV for $\theta_{\rm lab}=17^{\circ}$ (black histogram) and the $^{170}$Er($d$,$t$)$^{169}$Er reaction at 12 MeV for $\theta_{\rm lab}=90^{\circ}$ (red histogram). (d) The same reactions as (c) on the $^{174}$Yb target.}
\end{figure}

\begin{table}
\caption{\label{tab2} Comparison of normalization factors derived from the Niels Bohr Institute (NBI) data and the Munich (MLL) data.}
\newcommand\T{\rule{0pt}{3ex}}
\newcommand \B{\rule[-1.8ex]{0pt}{0pt}}
\begin{ruledtabular}
\begin{tabular}{cccc}
\B\T Isotope(s) & Reactions & Dataset &  $\overline{N}$  \\
\hline
\T All\footnote{All 21 cases used in the analyses shown in Table~\ref{tab1}} & ($d$,$p$) \& ($d$,$t$) & NBI & 1.18(15)  \\
\T $^{170}$Er \& $^{174}$Yb & ($d$,$p$) \& ($d$,$t$) & MLL & 1.08(6) \\
\T $^{170}$Er \& $^{174}$Yb & ($d$,$p$) \& ($p$,$d$) & MLL & 0.71(10) \\
\end{tabular}
\end{ruledtabular}
\end{table}

The product of the target thickness and spectrometer aperture were determined from elastic scattering yields of 12-MeV deuterons at $\theta_{\rm lab}=20^{\circ}$. Under these conditions, the cross section is estimated to be within 1\% of the Rutherford scattering cross section. The $^{170}$Er and $^{174}$Yb targets, isotopically enriched to 96.1\% and 95.8\% and supported on carbon backings of nominal thickness 20~$\mu$g/cm$^2$, were determined to have a thickness of 44(2)~$\mu$g/cm$^2$ and 50(3)~$\mu$g/cm$^2$, respectively.

Yields from the ($d$,$p$) and ($d$,$t$) reactions were also measured at 12~MeV to follow the previous NBI studies. For the ($d$,$p$) measurements, the cross sections were determined at angles of $\theta_{\rm lab}=60^{\circ}$ and 90$^{\circ}$, again used in the NBI studies, and an additional angle of 40$^{\circ}$. Similarly, with the ($d$,$t$) measurements, data were collected at angles of $\theta_{\rm lab}=60^{\circ}$ and 90$^{\circ}$, as well as $\theta_{\rm lab}=30^{\circ}$ for $^{170}$Er. The additional angles were added to guide choices in the optical-model parameterizations used in the analysis. The ($p$,$d$) reaction has more distinctive forward-peaked angular distributions and the ($d$,$p$) and ($d$,$t$) reactions. The ($p$,$d$) reaction yields were measured at $\theta_{\rm lab}=17^{\circ}$ and 38$^{\circ}$, which were estimated to be the peak angles for $\ell=1$ and 3 transfer cross sections. For the $^{174}$Yb($p$,$d$) reaction, only $\theta_{\rm lab}=17^{\circ}$ data were analyzed due to an incorrect field setting at $\theta_{\rm lab}=38^{\circ}$.

Example spectra are shown in Fig.~\ref{fig3} for each reaction. The spectra were calibrated using well-known states in the literature and the bands of interest are labeled, emphasizing the distinctive fingerprints of their energies and strengths. Due to the dispersion of the Q3D spectrometer, only a small range of excitation energy is probed in one magnet setting. The $Q$-value resolution achieved was $\sim$10~keV FWHM across the different reactions. The broad background features seen in the ($d$,$p$) reaction spectra are from the carbon backings and were also seen in the original NBI data, however, due to the dispersion of the Munich Q3D spectrograph, these appear as broad, weak peaks. The absolute cross sections, which are given in the Supplemental Material~\cite{supmat}, have estimated uncertainties of around 5\%.

The Munich (MLL) data for the ($d$,$p$) and ($d$,$t$) reactions were analyzed in the same way as that described above for the reanalysis of the NBI data. As with the NBI dataset, the results from the Munich experiment reveal consistent summed strengths (Eqn.~\ref{eqn2}) across the nuclei studied and the different Nilsson states. Table~\ref{tab2} summarizes the normalization factors derived from these analyses. The ($d$,$p$) and ($d$,$t$) datasets from NBI and MLL are in close agreement, with $\overline{N}=1.18(15)$ and 1.08(6), respectively. 

\begin{figure}
\includegraphics[scale=0.84]{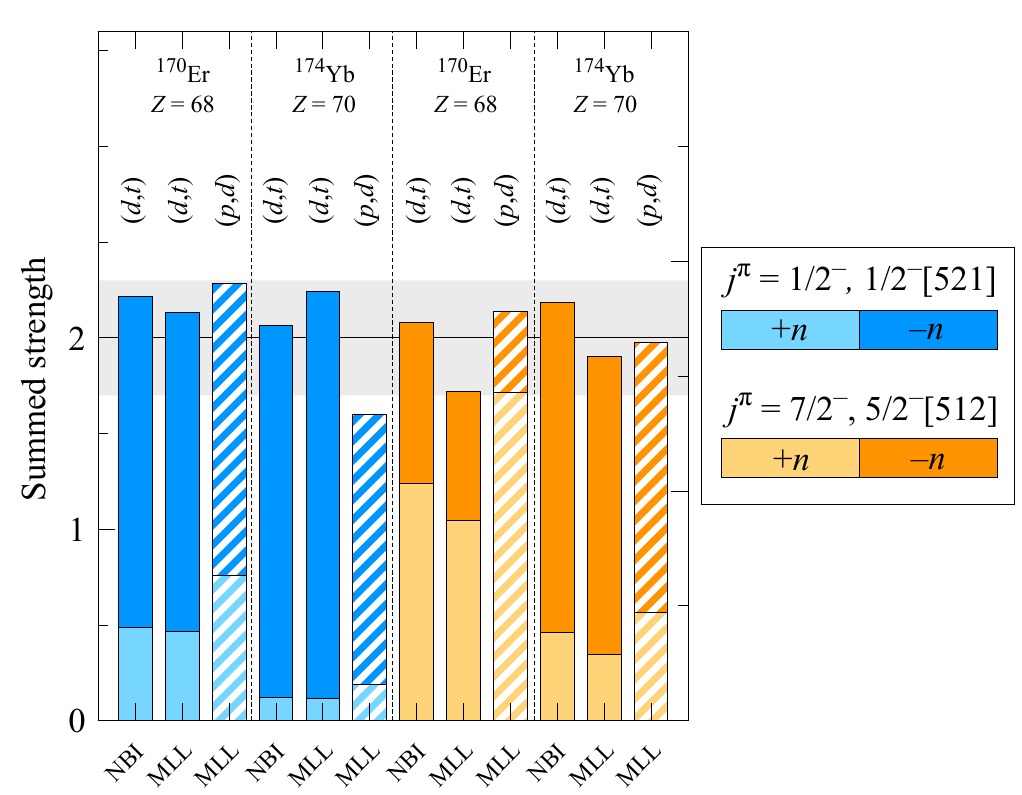}
\caption{\label{fig4} Information similar to that given in Fig.~\ref{fig1}, comparing the new MLL data to the NBI data, labeled by the source of the data and by reaction, with the sum being ($d$,$p$) [$+n$] and either ($d$,$t$) or ($p$,$d$) [$-n$] as indicated. The ($p$,$d$) data are also hatched to distinguish them from the ($d$,$t$) data.}
\end{figure}

In the case of the new ($p$,$d$) data, the same proton and deuteron optical-model parameterizations were used as for the ($d$,$p$) analysis. As with the analysis of the ($d$,$p$) and ($d$,$t$) reactions, good consistency is seen in the summed strengths determined from the ($d$,$p$) and ($p$,$d$) reactions, across the different isotopes and states of different angular momenta. The normalization for MLL ($d$,$p$) and ($p$,$d$) sums is lower, with $\overline{N}=0.71(10)$, which is a value in line with data from spherical systems. A comparison of the NBI and MLL datasets for ($d$,$p$), ($d$,$t$), and ($p$,$d$) data is show in Fig.~\ref{fig4}. 

The value of the normalization describes the degree to which single-particle motion is quenched. In these deformed nuclei, when the normalization value is determined from the summed strength derived from the ($d$,$p$) and ($d$,$t$) reactions at incident beam energies of 12~MeV, there is no apparent quenching, with $\overline{N}\sim1$. However, when determined from the neutron adding ($d$,$p$) reaction at 12~MeV, but with the neutron-removing strength calculated using yields from the ($p$,$d$) reaction at 18~MeV, the quenching factor is $\sim$0.7, which is a value consistent with expectations~\cite{Kramer01,Kay13}. As described above, the value of $\sim$1 determined from sums which make use of the ($d$,$t$) reaction is suspected to be due to a deficiency in the DWBA calculated cross sections for the ($d$,$t$) reaction.

\section{Summary}

In summary, we have studied how well single-nucleon transfer sum rules are obeyed in well-deformed nuclei, reanalyzing in a consistent way data published by the NBI group in the 1960s and 1970s. The data are for fifteen pairs of neutron-adding and removing cross sections for the $1/2^-$ members of the $1/2^-[521]$ band and six of the $j^{\pi}=7/2^-$ member of the $5/2^-[512]$ band, between neutron numbers $N=94-108$. Given the assumption that the states populated are pure Nilsson states, and un-fragmented by several mixing mechanisms known in deformed nuclei, remarkable consistency is found, when comparing equivalent sums across a broad range of nuclei. 

The value of the normalization required to satisfy the sum rules for the combined ($d$,$p$) and ($d$,$t$) reactions is found to be $\sim$1 instead of the $\sim$0.6 characteristic of ($d$,$p$) and ($p$,$d$) data in spherical nuclei. Analysis of new data from Munich, using the ($p$,$d$) reaction at above barrier energies as an alternative to the ($d$,$t$) reaction, yield a normalization around 0.7, more nearly consistent with the large body of data using various probes [though not including ($d$,$t$) data] from closed and near-closed shell spherical systems. The origin of the discrepancy is not immediately obvious. Two factors to be considered are a the treatment of deformation in reaction theory, and the modeling of the ($d$,$t$) reaction is likely deficient, which appears to be more sensitive to radii than ($d$,$p$) or ($p$,$d$) reactions.

In the era of radioactive ion beams, there has been a resurgence of interest in interpreting transfer reaction data for lighter nuclei using the Nilsson model~\cite{Macchiavelli17,Macchiavelli18,Macchiavelli20}. As data become more plentiful, exploiting the sum rules across wider ranges of isotopes will become increasingly important.

We are indebted to Filip Kondev and Augusto Macchiavelli for sharing their knowledge on these matters. We would to thank John Greene (Argonne) for making the targets for this measurement and the staff at the Maier-Leibnitz Laboratorium der Münchner Universitäten. This material is based upon work supported by the U.S. Department of Energy, Office of Science, Office of Nuclear Physics, under Contract Number DE-AC02-06CH11357, the UK Science and Technology Facilities Council under Grant Nos. ST/P004423/1 (Manchester) and ST/P005101/1 (West of Scotland), and the Deutsche Forschungsgemeinschaft Cluster of Excellence ``Origin and Structure of the Universe.''


\end{document}